\documentclass[aps,prb,twocolumn,showpacs,superscriptaddress]{revtex4-2}

\usepackage{graphicx}
\usepackage{amsmath}
\usepackage{multirow}
\usepackage{tabularx}
\usepackage{color}
\usepackage{float}
\usepackage[colorlinks,linkcolor=blue,anchorcolor=blue,citecolor=blue]{hyperref}

\begin{document}

\title{Chiral magnon splitting in altermagnetic CrSb from first principles}

\author{Yi-Fan Zhang}
\affiliation{Center for Neutron Science and Technology, Guangdong Provincial Key Laboratory of Magnetoelectric Physics and Devices, State Key Laboratory of Optoelectronic Materials and Technologies, School of Physics, Sun Yat-Sen University, Guangzhou 510275, China}

\author{Xiao-Sheng Ni}
\affiliation{Center for Neutron Science and Technology, Guangdong Provincial Key Laboratory of Magnetoelectric Physics and Devices, State Key Laboratory of Optoelectronic Materials and Technologies, School of Physics, Sun Yat-Sen University, Guangzhou 510275, China}

\author{Ke Chen}
\email{chenk35@mail.sysu.edu.cn}
\affiliation{Center for Neutron Science and Technology, Guangdong Provincial Key Laboratory of Magnetoelectric Physics and Devices, State Key Laboratory of Optoelectronic Materials and Technologies, School of Physics, Sun Yat-Sen University, Guangzhou 510275, China}

\author{Kun Cao}
\email{caok7@mail.sysu.edu.cn}
\affiliation{Center for Neutron Science and Technology, Guangdong Provincial Key Laboratory of Magnetoelectric Physics and Devices, State Key Laboratory of Optoelectronic Materials and Technologies, School of Physics, Sun Yat-Sen University, Guangzhou 510275, China}

\begin{abstract}
Altermagnetism has been proposed as a new type of magnetism, simultaneously exhibiting compensated spin moments in real space and spin-split electronic bands in reciprocal space. Alternating chiral magnon splitting is considered a unique feature of altermagnets. In this work, utilizing linear spin wave theory (LSWT), which is based on a localized spin picture and itinerant time-dependent density functional perturbation theory (TD-DFPT), we investigate the spin fluctuation spectra of altermagnetic CrSb. Along the L-$\Gamma$-L$^{\prime}$ path, the LSWT provides a chiral magnon splitting of up to 9 meV, located at high excitation energies around 140 meV, which is identified to be primarily driven by the splitting of two long-range exchange interactions, with exchange paths along the body diagonal lines of the unit cell. On the other hand, the more realistic TD-DFPT obtains more significant splitting of $\sim$ 30 meV at maximum. However, the splitting is severely smeared out due to strong Landau damping from the Stoner continuum, which may make it difficult to observe experimentally, e.g. through inelastic neutron scattering. We further provide a brief discussion on the connection between the Stoner excitations and the chiral magnon splitting.
      
\end{abstract}

\maketitle

\section{Introduction}\label{introduction}

Altermagnetism (AM) represents a newly proposed magnetic phase that bridges the gap between conventional ferromagnetism and antiferromagnetism~\cite{PhysRevX.12.031042,PhysRevX.12.040501,PhysRevX.12.040002}. In altermagnets, spins are arranged in a compensated antiparallel configuration akin to antiferromagnets, resulting in zero net magnetization. However, unlike antiferromagnets, AM systems exhibit spin splitting in momentum space due to broken crystalline symmetries. This spin splitting, reminiscent of ferromagnets, arises without requiring spin-orbit coupling, distinguishing AM from traditional magnetic systems. These characteristics give rise to unique electronic and spintronic properties that are advantageous for various technological applications, such as the anomalous Hall effect ~\cite{sciadv.aaz8809}, nonrelativistic spin-polarized currents ~\cite{PhysRevLett.126.127701,Naka2019,Shao2021,Ma2021,PhysRevB.103.125114}, spin torque effects ~\cite{PhysRevLett.128.197202,PhysRevLett.129.137201}, magnetic circular dichroism~\cite{PhysRevLett.132.176701}, and the piezomagnetic effect~\cite{PhysRevMaterials.8.L041402}. The discovery of altermagnetism has opened new research directions, particularly in developing magnetic memory and other next-generation spintronic technologies.

Symmetry analysis combined with first-principles calculations has identified several promising candidates for AM~\cite{GUO2023100991}. Among these, rutile RuO$_2$ and hexagonal MnTe have emerged as key prototypes due to their distinctive spin-split features, magnetic transport properties, and spin-torque effects~\cite{PhysRevLett.129.137201,PhysRevLett.130.216701,Feng2022,PhysRevLett.130.036702}. While much of the existing research has focused on the unique electronic structures of altermagnets, increasing attention has also been directed toward their spin dynamics. Using a phenomenological model, Steward \textit{et al.} examined spin excitations in altermagnets by focusing on the coupling between lattice dynamics and spin excitations, as revealed in the phonon spectra~\cite{PhysRevB.108.144418}. In RuO$_2$, spin fluctuations have been studied using linear spin wave theory (LSWT) and weak coupling theory, revealing alternating chirality splitting in magnon bands—distinct from the chirality-degenerate magnon bands typically found in conventional antiferromagnets~\cite{PhysRevLett.132.176701,PhysRevB.108.L100402}. Recently, Liu \textit{et al.} provided the first experimental evidence of alternating magnon splitting in MnTe through inelastic neutron scattering~\cite{PhysRevLett.133.156702}. The hexagonal NiAs-type compound CrSb (space group: P6$_3$/$mmc$) has also emerged as a strong candidate for AM~\cite{GUO2023100991}. Recent spin-integrated soft X-ray angle-resolved photoelectron spectroscopy (ARPES) experiments have observed band splitting in CrSb thin films~\cite{Reimers2024,PhysRevLett.133.206401}. With a Néel temperature close to 700 K~\cite{PhysRev.85.365,PhysRev.129.2008} and a predicted spin splitting of approximately 1 eV~\cite{GUO2023100991}, CrSb stands out among predicted AM candidates, warranting further investigation.


In this work, we investigate spin fluctuations in CrSb using LSWT and a fully \textit{ab initio} approach, based on the analysis of the dynamic spin susceptibility within time-dependent density functional perturbation theory (TD-DFPT). The spin fluctuation spectra are calculated along high-symmetry lines $\Gamma$-A-L-$\Gamma$-L$^{\prime}$ in the Brillouin zone (BZ). The LSWT reveals alternating magnon splitting, with a maximum amplitude of 9 meV along the L-$\Gamma$-L$^{\prime}$ direction. This splitting is primarily attributed to the difference between the long-range eleventh and twelfth nearest-neighbor exchange interactions, $J_{11}$ and $J_{12}$. In contrast, the TD-DFPT, which adopts an itinerant spin model, predicts significantly larger magnon splitting, with a maximum amplitude of approximately 34 meV along the same path. However, the most pronounced splitting occurs at high excitation energies around 150 meV, where it is substantially smeared by Landau damping. This damping effect may pose challenges for experimental observation using techniques such as inelastic neutron scattering. Finally, we explore the relationship between the electronic band structure and chiral magnon splitting.

\section{Formalism}\label{formalism}

Our \textit{ab initio} calculations were performed using the Quantum Espresso suite~\cite{giannozzi2009quantum}. For the exchange and correlation functional, we employed the local density approximation (LDA)~\cite{perdew1981self}, along with corresponding norm-conserving pseudopotentials from the 'PSlibrary 0.3.1' repository~\cite{dal2014pseudopotentials}. A plane-wave energy cutoff of 80 Ry was used, and Gaussian smearing was applied for BZ integration in the presence of a Fermi surface, with a smearing width of 10 mRy. The crystal structure of CrSb was obtained from the MAGNDATA database~\cite{annurev}, with experimental lattice parameters \( a = 4.127 \, \text{\AA} \) and \( c = 5.451 \, \text{\AA} \). A \( 15 \times 15 \times 12 \) grid of \( \mathbf{k} \)-points was used for both the ground state calculations and the solution of the Sternheimer equation. For our LSWT calculations, we used the SpinW software package~\cite{toth2015linear}, where \textbf{S} was set to \( \frac{3}{2} \). The TB2J code~\cite{HE2021107938} was employed to compute the magnetic exchange interactions, which were derived using maximally localized Wannier functions (MLWFs) generated by the Wannier90 code~\cite{Pizzi2020}.

We now briefly describe the formalism of TD-DFPT we employed. Unless otherwise stated, atomic units (Rydberg units) are used throughout. TD-DFPT follows a two-step procedure to calculate the generalized spin susceptibility~\cite{buczek2011different, savrasov1998linear, rousseau2012efficient, karlsson2000many, csacsiouglu2010wannier}. Firstly, the Kohn-Sham susceptibility is given by
\begin{equation} 
	\begin{aligned}
		\chi_{\mathrm{KS}}^{i j}\left(\mathbf{r}, \mathbf{r}^{\prime}, \omega\right)=& \frac{1}{N_{\mathbf{k}}^{2}} \sum_{n m, \mathbf{k}, \mathbf{q}} \frac{f_{n \mathbf{k}}-f_{m \mathbf{k}+\mathbf{q}}}{\epsilon_{n \mathbf{k}}-\epsilon_{m \mathbf{k}+\mathbf{q}}+\omega} \\
		& \times \vec{\psi}_{n \mathbf{k}}^{\dagger}(\mathbf{r}) \sigma^{i} \vec{\psi}_{m \mathbf{k}+\mathbf{q}}(\mathbf{r}) \vec{\psi}_{m \mathbf{k}+\mathbf{q}}^{\dagger}\left(\mathbf{r}^{\prime}\right) \sigma^{j} \vec{\psi}_{n \mathbf{k}}\left(\mathbf{r}^{\prime}\right)
	\end{aligned} \label{eq:1} 
\end{equation}
where $i,j$ = 0, ..., 3, 0 denotes charge degree of freedom and 1, 2, 3 represent three Cartesian directions $x, y, z$. $ N_{\mathbf{k}} $ means the number of $\mathbf{k}$ points used to discretize the first BZ (we assume a uniform sampling). $f_{n \mathbf{k}}$ and $f_{m \mathbf{k}+\mathbf{q}}$ are occupation numbers. $\vec{\psi}_{n \mathbf{k}}$ is a Kohn-Sham two-spinor eigenfunction with wave vector $\mathbf{k}$, band index $\textit{n}$, and energy $\epsilon_{n \mathbf{k}}$, and the asterisk indicates complex conjugation. $\sigma$ is the usual 2 $\times$ 2 Pauli matrices, with $\sigma^0=\hat{I}$. The Hartree and exchange-correlation potentials are modified by the induced charge and magnetization densities, which are described by the Kohn-Sham susceptibility, introducing a self-consistency issue. These densities affect the effective fields and are simultaneously induced by them. The second step of this formalism addresses the self-consistency~\cite{gross1985local}.
\begin{equation} 
	\begin{aligned}
		&\chi^{i j}\left(\mathbf{r} t, \mathbf{r}^{\prime} t^{\prime}\right) \\
		&=\chi_{\mathrm{KS}}^{i j}\left(\mathbf{r} t, \mathbf{r}^{\prime} t^{\prime}\right)+\sum_{k l} \int d\left(\mathbf{r}_{1} t_{1}\right) d\left(\mathbf{r}_{2} t_{2}\right) \\
		&\quad \times \chi_{\mathrm{KS}}^{i k}\left(\mathbf{r} t, \mathbf{r}_{1} t_{1}\right)\left[f_{x c}^{k l}\left(\mathbf{r}_{1} t_{1}, \mathbf{r}_{2} t_{2}\right)+\frac{2 \delta_{k 0} \delta_{10} \delta\left(t_{1}-t_{2}\right)}{\left|\mathbf{r}_{1}-\mathbf{r}_{2}\right|}\right] \\
		&\quad \times \chi^{l j}\left(\mathbf{r}_{2} t_{2}, \mathbf{r}^{\prime} t^{\prime}\right)
	\end{aligned} \label{eq:2} 
\end{equation}
The frequency-independent functional derivative of the LSDA exchange-correlation potential is used to approximate the exchange-correlation kernel $f_{x c}^{i j}$~\cite{buczek2011different,gross1985local,von1972local,callaway1975transverse},
\begin{equation}
	\begin{aligned}
		f_{x c}^{i j}\left(\mathbf{r} t, \mathbf{r}^{\prime} t^{\prime}\right)=\frac{\delta^{2} E_{x c}}{\delta \rho^{i}(\mathbf{r}) \rho^{j}\left(\mathbf{r}^{\prime}\right)} \delta\left(\mathbf{r}-\mathbf{r}^{\prime}\right) \delta\left(t-t^{\prime}\right)
	\end{aligned} \label{eq:3} 
\end{equation}
The main drawback of this procedure is the need to compute unoccupied Kohn-Sham states in order to evaluate $\chi^{KS}$ through Eq.~(\ref{eq:2}). Convergence with respect to these unoccupied bands is often very slow. An alternative approach, based on solving the Sternheimer equation, allows us to bypass this bottleneck by computing only the occupied Kohn-Sham states. The results obtained using this method are consistent with those from the first approach.

The Sternheimer equation reads
\begin{equation} 
	\begin{aligned}
		\left(\hat{H}-i\frac{\partial}{\partial t} \hat{I}\right)\delta \vec{\psi}_{n \mathbf{k}}(\mathbf{r}, t)=-\left(1-\hat{P}_{\mathrm{occ}}\right) \delta \hat{V}_{\mathrm{scf}}(\mathbf{r},t) \vec{\psi}_{n \mathbf{k}}(\mathbf{r})
	\end{aligned} \label{eq:4} 
\end{equation}
where $\hat{H}$ denotes the unperturbed Kohn-Sham Hamiltonian (corresponding to the term in square brackets), $\delta \vec{\psi}_{n \mathbf{k}}$ represents the first-order change in the spinor wave function, and $\delta\hat{V}_{\mathrm{scf}}$ is the first-order variation of the Kohn-Sham potential. The operator $\hat{P}_{\mathrm{occ}}$ is the projector onto the manifold of unoccupied Kohn-Sham states. In this scheme, a self-consistent ground-state calculation is first performed to obtain the ground-state electron density and Kohn-Sham wavefunctions. An external plane-wave magnetic field is then applied to the right-hand side of Eq.\ref{eq:4}, which is solved self-consistently to obtain the first-order response of the spin density, and thus the generalized spin susceptibility. Our implementation is based on the linear-response modules of the QUANTUM ESPRESSO materials simulation suite~\cite{giannozzi2009quantum}. More details of this formalism can be found in Ref.~\cite{cao2018ab}.

\begin{figure}[H]
	\centering
	\includegraphics[scale=0.22]{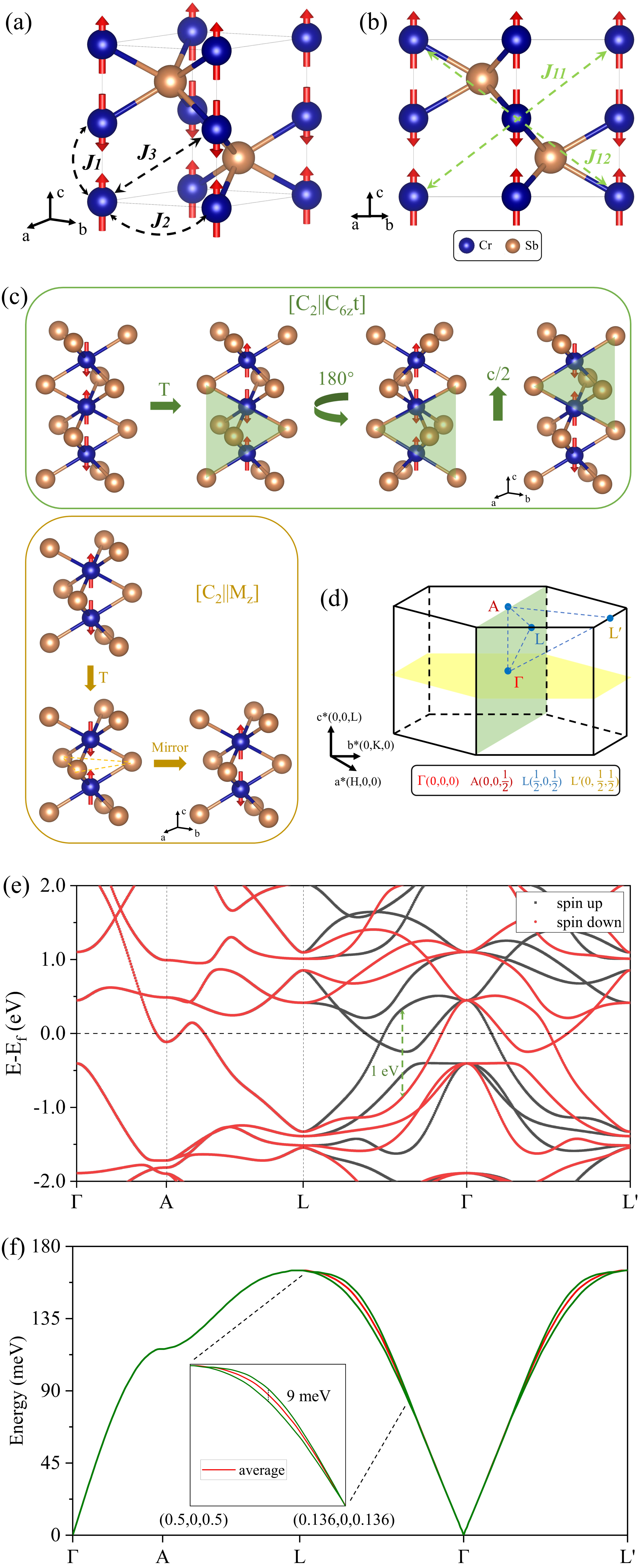}
	\caption{(a),(b) Crystal and magnetic structures of CrSb. Blue spheres represent Cr atoms and Golden spheres represent Sb atoms. Arrows denote magnetic moments of Cr atoms. The black dashed lines indicate the paths of the first three nearest-neighbor exchange interactions. The $J_{11}$ and $J_{12}$ are indicated by green dashed lines. (c),(d) 3D BZ of CrSb with the selected $\Gamma$-A-L-$\Gamma$-L$^{\prime}$ high-symmetry paths. The position of each high-symmetry point is given in the lower panel. The green (yellow) plane refers to the nodal surface protected by \([C_2 \parallel C_{6z}t]\) (\([C_2 \parallel M_z]\)). (e) Calculated electronic band structures of CrSb. The band splitting up to 1 eV is indicated by the green line. (f) Calculated spin wave spectra using LSWT along $\Gamma$-A-L-$\Gamma$-L$^{\prime}$ directions, the red solid line represents the LSWT result obtained by averaging $J_{11}$ and $J_{12}$.}
	\label{fig1} 
\end{figure}

\section{Results}\label{results}

As shown in Fig.~\ref{fig1}(a), our \textit{ab initio} calculations predict an A-type collinear antiferromagnetic (AFM) ground state, with the Néel vector oriented along the \( c \)-axis and a magnetic moment of 2.70 \(\mu_B\)/Cr, in excellent agreement with experimental values of 2.7 \(\mu_B\)/Cr~\cite{ITO2007e558,PhysRev.85.365}. Each Cr atom, situated at the center of an octahedron formed by six Sb atoms, forms two distinct sublattices with antiparallel spins. The full symmetry of CrSb can be described using the spin-group formalism~\cite{doi:10.1098/rspa.1966.0211, LITVIN1974538, Litvin:a14103}, which generalizes conventional magnetic groups. In this framework, symmetry operations are represented as pairs \([R_i \parallel R_j]\), where the operations on the left act on the spin space, and those on the right act on real space. Fig.~\ref{fig1}(c) illustrates the relation between the spin sublattices in CrSb via nonrelativistic spin-group operations, namely \([C_2 \parallel C_{6z}t]\) and \([C_2 \parallel M_z]\). Here, \(C_2\) denotes a 180° spin-space rotation perpendicular to the spins. On the real-space side, \(C_{6z}t\) represents a combined six-fold rotation and translation operation, while \(M_z\) is a mirror operation. These symmetry operations protect the spin degeneracy in four nodal planes, as shown in Fig.~\ref{fig1}(d), in addition to the BZ boundaries. Our calculated electronic band structure, displayed in Fig.~\ref{fig1}(e), reveals significant spin splitting away from the high-symmetry planes, particularly along the L-$\Gamma$-L$^{\prime}$ direction. The calculated spin splitting exceeds 1 eV, which is in good agreement with recent theoretical~\cite{GUO2023100991} and experimental findings~\cite{PhysRevLett.133.206401}. These results underscore the interplay between spin-group symmetry and electronic band structure in CrSb.

\subsection{LSWT}\label{LSWT}

We first map the magnetic energy of CrSb to the following Heisenberg model,

\begin{equation} 
	\begin{aligned}
		\textit{H} = J_{ij}\sum_{i<j}\textbf{S}_i\ \cdot\ \textbf{S}_j 
	\end{aligned}
\end{equation}	

where \textbf{S}$_i$ corresponds to the local spin of Cr ions. As shown in Fig.~\ref{fig1}(a) and Fig.~\ref{fig1}(b), we consider exchange interactions up to the twelfth-nearest-neighbor $J_{12}$. 
\begin{table*}
	\caption{\label{tab:table1} Heisenberg exchange couplings for the LSWT.}
	\begin{ruledtabular}
		\begin{tabular*}{\linewidth}{@{}ccccccccccccc@{}}
			CrSb&$J_{1}$&$J_{2}$&$J_{3}$&$J_{4}$&$J_{5}$&$J_{6}$&$J_{7}$&$J_{8}$&$J_{9}$&$J_{10}$&$J_{11}$&$J_{12}$\\ \hline
			$J_{ij}$ (meV)&25.76&-4.56&3.02&0.75&-0.62&0.75&-0.38&-0.12&-0.12&0.09&0.42&1.10\\
			Bond length (\AA)&2.725&4.127&4.946&5.451&6.837&7.148&7.65&8.176&8.254&8.692&8.989&8.989\\
		\end{tabular*}
	\end{ruledtabular}
\end{table*}

The calculated exchange interactions \(J\) are listed in Table~\ref{tab:table1}. Among these, the first three interactions, \(J_1\), \(J_2\), and \(J_3\), are an order of magnitude larger than the others, which primarily drive a ferromagnetic alignment of the magnetic moments within each Cr layer and AFM coupling between adjacent layers. Notably, compared to other altermagnetic candidates, CrSb exhibits a particularly robust magnetic ordering, characterized by a high Néel temperature \(T_N\) $\sim$ 705 K. Classical Monte Carlo simulations were performed~\cite{Markov,PhysRevLett.103.257201} using the calculated exchange interactions. The simulated Néel temperature \(T_N\) is approximately 685 K, which is in excellent agreement with the experimental value~\cite{PhysRev.85.365,PhysRev.129.2008}, validating the reliability of the calculated exchange interactions.

The calculated spin wave spectra, obtained using LSWT, along the high-symmetry $\Gamma$-A-L-$\Gamma$-L$^{\prime}$ directions in the BZ, are shown as green lines in Fig.~\ref{fig1}(f). The high-symmetry $\Gamma$-A-L-$\Gamma$-L$^{\prime}$ paths and $K$ points in the BZ are marked in Fig.~\ref{fig1}(d). Linear magnon dispersion can be observed around the $\Gamma$ point, exhibiting behavior similar to that of conventional antiferromagnets, with excitation energy reaching approximately 165 meV at the L point. On the $\Gamma$-A-L path, only one degenerate magnon band is obtained. However, along the L-$\Gamma$-L' path, the degeneracy is lifted, with a maximum splitting of 9 meV. In the next section, we will show that the split bands correspond to opposite chiralities, which closely mirrors the spin splitting of electronic band structures observed along the same paths.

From the perspective of the Heisenberg model, we find that the splitting of the magnons is primarily driven by the splitting between two long-range exchange interactions \(J_{11}\) and \(J_{12}\), which both have the same bond length of 8.989 $\rm \AA$. The corresponding exchange paths lie along the body diagonal directions of the unit cell, with the path of \(J_{12}\) (along [1 -1 -1]) crossing the Sb atoms, unlike \(J_{11}\) (along [1 -1 1]), as shown in Fig.~\ref{fig1}(b). Similarly, in previous theoretical studies on RuO$_2$ and MnTe, two long-range exchange interactions with equal bond lengths but different symmetries were also identified as the leading cause of magnon splitting~\cite{PhysRevLett.131.256703,PhysRevLett.133.156702}. Interestingly, unlike RuO$_{2}$ and MnTe, where the two exchange interactions are found of opposite signs~\cite{PhysRevLett.131.256703,PhysRevLett.133.156702}, our calculations indicate that in CrSb, \(J_{11}\) and \(J_{12}\) have the same sign. Nevertheless, a small difference in their magnitude is sufficient to cause significant magnon splitting. To verify, we further calculate the linear spin wave by artificially setting \(J_{11}\) and \(J_{12}\) to the average of their true values. In this case, the magnon splitting vanishes, as shown in Fig.~\ref{fig1}(f), thereby confirming that the observed magnon splitting arises from the disparity in their magnitudes. 

\subsection{TD-DFPT}\label{TDDFPT}

In the context of linear response theory, the complete information about the spin-wave spectra and the Stoner continuum can be obtained by calculating the spin susceptibility \(\chi(\mathbf{q}, \omega)\) and the Kohn-Sham non-interacting susceptibility \(\chi_{\text{KS}}(\mathbf{q}, \omega)\). The magnon excitation energies correspond to the poles of \(\text{Im}\,\chi(\mathbf{q}, \omega)\). In contrast, the nonzero region of \(\text{Im}\,\chi_{\text{KS}}(\mathbf{q}, \omega)\) in the \((\mathbf{q}, \omega)\)-plane defines the Stoner excitation continuum. For Stoner excitations, scattering processes involving positive chirality, described by \(\text{Im}\chi^{+-}_{\text{KS}}\), correspond to particle-hole excitations with spin \(\textbf{S} = +1\), where momentum \(\mathbf{q}\) transfers from a spin-down electron to a spin-up electron. Conversely, scattering processes with negative chirality, described by \(\text{Im}\chi^{-+}_{\text{KS}}\), involve particle-hole excitations with spin \(\textbf{S} = -1\), where momentum \(\mathbf{q}\) transfers from a spin-up electron to a spin-down electron. While Stoner excitations correspond to particle-hole transitions, magnons can be regarded as collective spin-wave excitations arising from similar processes. In this context, \(\text{Im}\chi^{+-}\) and \(\text{Im}\chi^{-+}\) describe magnons with positive and negative chirality, respectively.

In Sec.~\ref{formalism}, we present the formalism for calculating \(\chi^{ij}(\mathbf{r}, \mathbf{r}^{\prime}, t)\). By applying a Fourier transform to this formalism, the spin susceptibility is expressed in the frequency and momentum space:

\begin{equation} 
    \begin{aligned}
		\chi^{+-}(\mathbf{q}, \omega)=& \int d t e^{i(\omega+i \eta) t} \frac{1}{\Omega} \int d \mathbf{r} \int d \mathbf{r}^{\prime} \\
		& \times e^{-i\mathbf{q} \cdot \mathbf{r}} \chi^{+-}\left(\mathbf{r}, \mathbf{r}^{\prime}, t\right) e^{i\mathbf{q} \cdot \mathbf{r}^{\prime}}
    \end{aligned}
\end{equation}	
where $\mathbf{q}$ represents the wave vector confined to the first BZ. $\eta$ represents a positive infinitesimal. $\chi^{-+}$ can be written as $\chi^{-+} = \chi^{11} +i\chi^{12}$.
\begin{figure*}[!htb]
	\centering
	\includegraphics[scale=0.3]{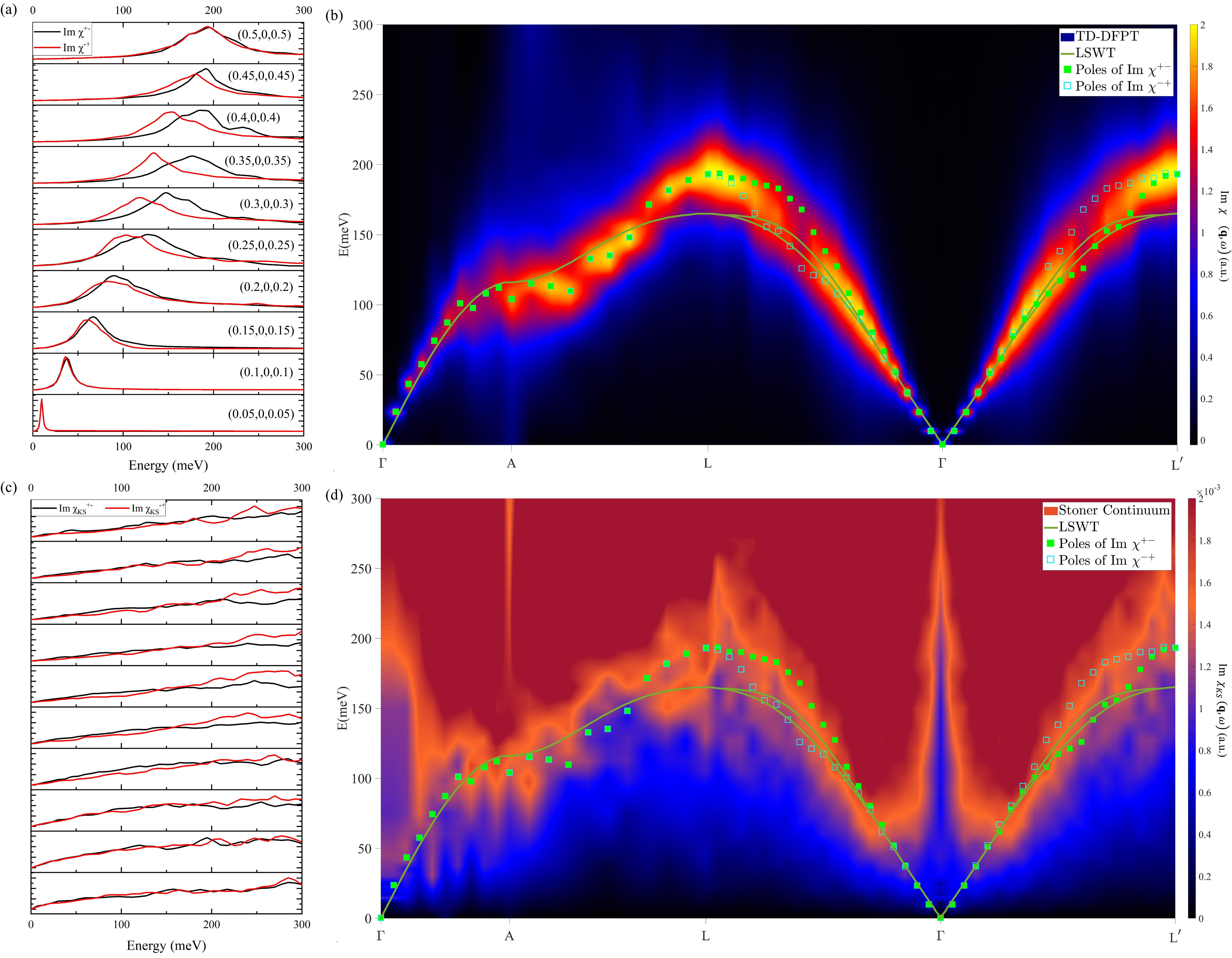}
	\caption{(a) Calculated Im$\chi^{+-}(\mathbf{q}, \omega)$ and Im$\chi^{-+}(\mathbf{q}, \omega)$ along $\Gamma$-L direction, the position of each $\mathbf{q}$ is indicated on the right side of the curves. (b) Spin fluctuation spectra of CrSb along the high-symmetry directions below 300 meV. The color map represents the calculated (Im$\chi^{+-}(\mathbf{q}, \omega)$+Im$\chi^{-+}(\mathbf{q}, \omega)$)/2. The LSWT results are shown as green lines. The green solid dots correspond to the poles of Im$\chi^{+-}$, and the cyan hollow dots correspond to the poles of Im$\chi^{-+}$. (c) Calculated Im$\chi^{+-}(\mathbf{q}, \omega)$ and Im$\chi^{-+}(\mathbf{q}, \omega)$ along $\Gamma$-L direction, (d) Calculated Stoner excitation spectra (Im$\chi_{K S}^{+-}(\mathbf{q}, \omega)$+Im$\chi_{K S}^{-+}(\mathbf{q}, \omega)$)/2.}
	\label{fig2} 
\end{figure*}

Using TD-DFPT, we calculate the transverse spin susceptibilities \(\text{Im}\,\chi^{+-}(\mathbf{q}, \omega)\) and \(\text{Im}\,\chi^{-+}(\mathbf{q}, \omega)\) for wave vectors \(\mathbf{q}\) along \(\Gamma\)-A-L-\(\Gamma\)-L\(^\prime\), with energies up to 300 meV. The average transverse spin susceptibility, (\(\text{Im}\,\chi^{+-}(\mathbf{q}, \omega) + \text{Im}\,\chi^{-+}(\mathbf{q}, \omega)\))/2, is shown as a color map in Fig.~\ref{fig2}(b), where brighter regions correspond to stronger intensity. The corresponding Stoner continuum is displayed in Fig.~\ref{fig2}(d). The peak positions of \(\text{Im}\,\chi^{+-}(\mathbf{q}, \omega)\) and \(\text{Im}\,\chi^{-+}(\mathbf{q}, \omega)\), representing the excitation energies of positive and negative chiral magnons, are marked for comparison with LSWT results. Well-defined magnon excitations are observed over the entire sampled high-symmetry path. At high excitation energies, typically above 80 meV, the magnons experience strong Landau damping due to interactions with dense Stoner continuum, as shown in Fig.~\ref{fig2}(d). Along the $\Gamma$-A-L path, the TD-DFPT results show chirally degenerate magnons with linear dispersion near the $\Gamma$ point, consistent with LSWT predictions. 
\begin{figure*}[!htb]
	\centering
	\includegraphics[scale=0.55]{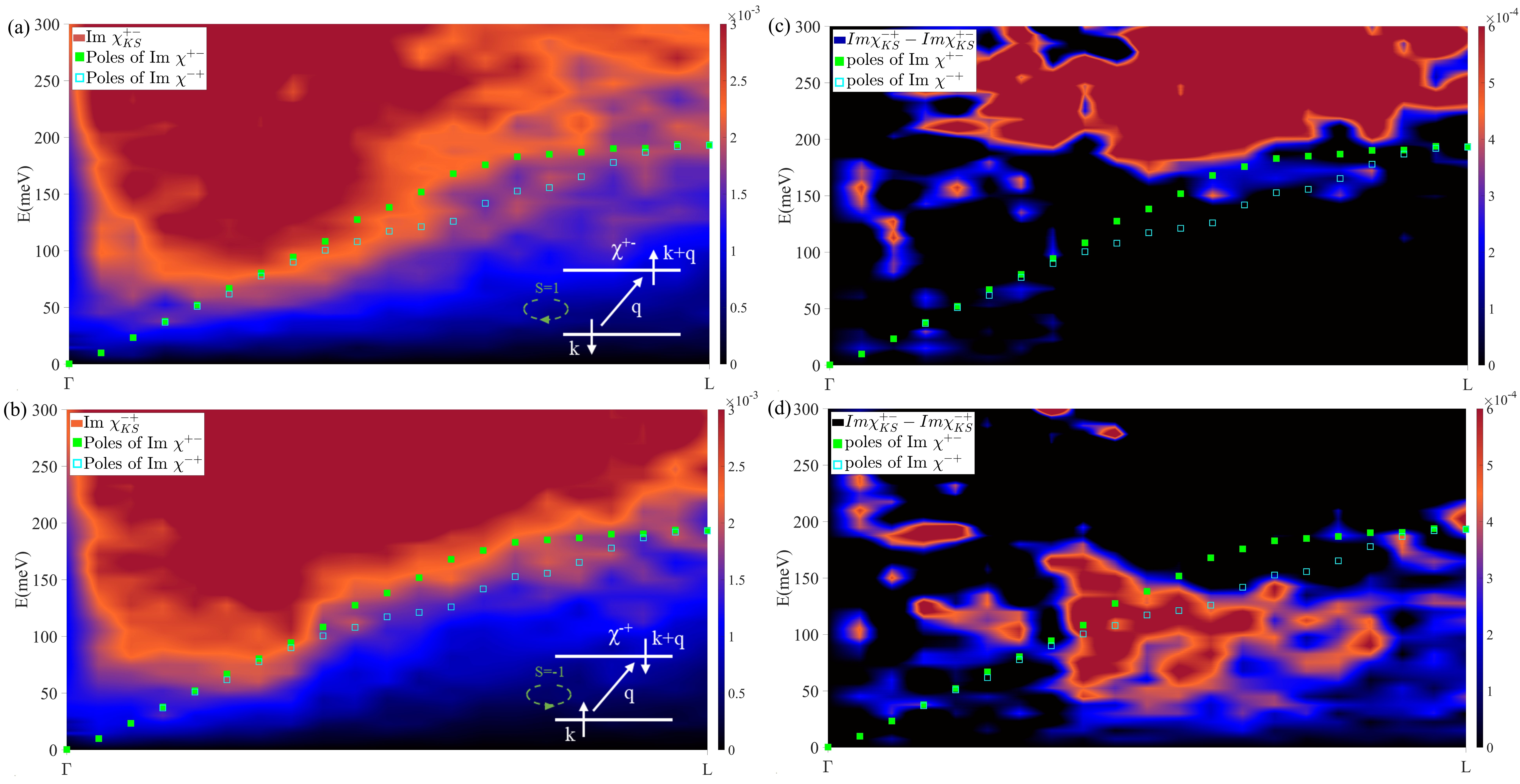}
	\caption{Stoner continuum of CrSb along the high-symmetry directions below 300 meV. The color map represents the (a) calculated Im$\chi_{K S}^{+-}(\mathbf{q}, \omega)$, (b) calculated Im$\chi_{K S}^{-+}(\mathbf{q}, \omega)$, (c) Im$\chi^{-+}(\mathbf{q}, \omega)$-Im$\chi^{+-}(\mathbf{q}, \omega)$, (d)  Im$\chi^{+-}(\mathbf{q}, \omega)$-Im$\chi^{-+}(\mathbf{q}, \omega)$. The green solid (cyan hollow) dots correspond to the poles of Im$\chi^{+-}$ (Im$\chi^{-+}$).}
	\label{fig3} 
\end{figure*}
%

Along the \(\Gamma\)-L path, as shown in Fig.~\ref{fig2}(b) and highlighted in selected constant-\(\mathbf{q}\) cuts in Fig.~\ref{fig2}(a),  the \(\text{Im}\,\chi^{+-}\) and \(\text{Im}\,\chi^{-+}\) remain nearly identical at the \(\Gamma\) and L points within numerical precision, with splitting becoming apparent away from these points. When the excitation energy reaches approximately 80 meV, noticeable differences begin to emerge between \(\text{Im}\,\chi^{+-}(\mathbf{q}, \omega)\) and \(\text{Im}\,\chi^{-+}(\mathbf{q}, \omega)\) with the maximum splitting reaching \(\sim\)30 meV around \(\mathbf{q} = \left(0.3, 0, 0.3\right)\), which is substantially larger than the 9 meV predicted by the LSWT.  Furthermore, the high-energy excitations around the L point are significantly higher than those predicted by LSWT, as commonly observed in itinerant magnetic systems such as elemental Fe, Co, and Ni~\cite{buczek2011different}, highlighting the strong renormalization effect of the Stoner continuum on high-energy excitations.

Although the magnon splitting from TD-DFPT is significantly larger than the 2 meV splitting observed recently in MnTe~\cite{PhysRevLett.133.156702}, it is strongly smeared by Landau damping, which may pose challenges for experimental detection, such as through inelastic neutron scattering. This damping is absent at low energies in semiconducting MnTe due to the presence of an energy gap, which removes the barrier for clearer experimental observation. Along the \(\Gamma\)-L path, the energy of the positive chiral magnons, as indicated by the peak positions of \(\text{Im}\,\chi^{+-}(\mathbf{q}, \omega)\), is consistently higher than that of the negative chiral magnons, which reverses along the \(\Gamma\)-L\(^\prime\) path. The alternating sign of the chirality splitting is a hallmark of altermagnets and has no counterpart in conventional ferromagnets or antiferromagnets. This phenomenon has also been observed in prior theoretical studies of magnons in the altermagnetic compound RuO$_2$~\cite{PhysRevLett.131.256703}.

 

Microscopically, the splitting of altermagnetic magnons can be traced back to the spin-split band structures in the same region of the BZ. The band splitting introduces a deviation between Im$\chi_{KS}^{-+}$ and Im$\chi_{KS}^{+-}$, partly due to differences in the denominator $\epsilon_{n \mathbf{k}} - \epsilon_{m \mathbf{k}+\mathbf{q}}$ of Eq.~(\ref{eq:1}) for the two excitation channels with opposite chiralities. This deviation propagates to $\chi^{-+}$ and $\chi^{+-}$, which are the dynamically screened counterparts of $\chi_{KS}^{-+}$ and $\chi_{KS}^{+-}$. However, as shown in the the comparison of Fig.~\ref{fig2}(a) and Fig.~\ref{fig2}(c), a direct and simple connection between the splitting of Stoner excitations and magnons cannot be established, as the dynamical screening involves additional complexities beyond the bare effects of the band structures. Despite this, we still plot the calculated $\chi_{KS}^{-+}$ and $\chi_{KS}^{+-}$ (see Fig.~\ref{fig3}(a) and Fig.~\ref{fig3}(b)), together with their difference in Fig.~\ref{fig3}(c) and Fig.~\ref{fig3}(d). Although the splitting between $\chi_{KS}^{-+}$ and $\chi_{KS}^{+-}$ is not significant, it is evident that the $\chi_{KS}^{-+}$ expels the corresponding $\chi^{-+}$, as $\chi_{KS}^{-+}$ weakens $\chi^{-+}$ through Landau damping. Consequently, as shown in Fig.~\ref{fig3}(c), in the high energy region, $\chi_{KS}^{-+}$ is stronger than $\chi_{KS}^{+-}$, as a result, the $\chi^{-+}$ is pushed down below the $\chi^{+-}$. In synergy, the stronger presence of the $\chi_{KS}^{+-}$ in the low energy region helps to push the $\chi^{+-}$ up (see Fig.~\ref{fig3}(d)).



\section{Summary}\label{summary}    

In summary, we investigate spin fluctuations in CrSb using LSWT and a fully \textit{ab initio} approach, based on the TD-DFPT. The spin fluctuation spectra are calculated along high-symmetry paths $\Gamma$-A-L-$\Gamma$-L$^{\prime}$ in the BZ. The LSWT reveals alternating magnon splitting, with a maximum amplitude of 9 meV along the L-$\Gamma$-L$^{\prime}$ direction, which is primarily attributed to the difference between the long-range eleventh and twelfth nearest-neighbor exchange interactions. In contrast, the TD-DFPT predicts significantly larger magnon splitting, with a maximum amplitude of approximately 34 meV. However, the most pronounced splitting occurs at high excitation energies around 150 meV, where it is substantially smeared by Landau damping. This damping effect may pose challenges for experimental observation using techniques such as inelastic neutron scattering. Finally, we explore the relationship between the electronic band structure and chiral magnon splitting.

%

\end{document}